\documentclass[twocolumn,showpacs,preprintnumbers,amsmath,amssymb]{revtex4}

\usepackage{graphicx}
\usepackage{dcolumn}
\usepackage{bm}
\usepackage{feynmp}

\newcommand{\setval}{\fmfset{wiggly_len}{1.5mm}\fmfset{arrow_len}{1.5mm}\fmfset{arrow_ang}{13}\fmfset{dash_len}{1.5mm}\fmfpen{0.125mm}\fmfset{dot_size}{1thick}}

\begin{document}

\preprint{APS/123-QED}

\title{High-Order Variational Perturbation Theory for the Free Energy}

\author{Florian Weissbach}
\email{florian.weissbach@physik.fu-berlin.de}
\author{Axel Pelster}
\email{pelster@physik.fu-berlin.de}
\author{Bodo Hamprecht}
\email{bodo.hamprecht@physik.fu-berlin.de}
\affiliation{%
Freie Universit\"at Berlin, Institut f\"ur Theoretische Physik,
Arnimallee 14, 14195 Berlin, Germany}%

\date{\today}

\begin{abstract}
In this paper we introduce a generalization to the algebraic 
Bender-Wu recursion
relation for the eigenvalues and the eigenfunctions of the anharmonic
oscillator. We extend this well known formalism to the time-dependent
quantum statistical Schr\"odinger equation, thus obtaining
the imaginary-time evolution amplitude by solving a recursive set
of ordinary differential equations.
This approach enables us to evaluate global and local quantum statistical
quantities of the anharmonic oscillator
to much higher orders than by evaluating
Feynman diagrams. We probe our perturbative results by deriving a perturbative
expression for the free energy which is then subject to variational
perturbation theory as developed by Kleinert, yielding convergent
results for the free energy for all values of the coupling strength.
\end{abstract}


\maketitle
\section{Introduction}
\label{sec1}
Most physical problems can only be solved by approximation
methods.
One of them is perturbation theory which yields weak-coupling
expansions.
Unfortunately, they often do not converge.\\
The ground state energy of the
anharmonic oscillator is the simplest example where this
phenomenon can be studied. Algebraic recursion relations \`{a} la Bender and Wu
\cite{Bender/Wu} yield perturbation series for
the eigenvalues (energy) and eigenfunctions (wave functions)
of the time-independent
Schr\"odinger equation up to arbitrarily high orders. 
In Ref.~\cite{exp} the calculation was extended to
250th order.
The Bender-Wu recursion relation yields a power series for
the anharmonic part of the wave function both in the
coupling strength $g$ and in the coordinate $x$. The power
series in $x$ can be cut off naturally by comparing
the recursive results with those obtained from
evaluating Feynman diagrams.
The resulting weak-coupling series for the ground state energy does
not converge for any value of the coupling strength.
This paper deals with both problems:
Obtaining high-order perturbation expressions and making them
converge for all values of the coupling strength. 
It is organized as follows:\\
In Sec.~\ref{sec2} we perturbatively evaluate the path integral representation
for the imaginary-time evolution amplitude of the anharmonic oscillator
by means of a generalized Wick's theorem \cite{weissbach,bachmann}.
In Sec.~\ref{sec3} we represent the first-order results diagrammatically.
Doing so, we demonstrate that the algebraic computational cost is
very high for the diagrammatic approach. We also obtain a
cross check for the results which are derived 
from a differential recursion relation in the subsequent
Sec.~\ref{sec4}. In order to cut down on the algebraic computational 
cost we introduce a strategy to exploit the
symmetry property of the imaginary-time evolution amplitude in 
Sec.~\ref{sec5},
thus laying the foundation for our high-order results.
In Sec.~\ref{sec6} we combine the resulting algebraic 
recursion relation with the
original differential recursion relation, 
thus generalizing the Bender-Wu approach
\cite{Bender/Wu}.
From our perturbative results for the imaginary-time evolution amplitude
we then gain a perturbation 
expression for the free energy of the anharmonic oscillator
in Sec.~\ref{sec7} which we check again diagrammatically
in Sec.~\ref{sec8}.
The perturbative results are then re-summed 
in Sec.~\ref{sec9} by means of
variational perturbation theory \cite{Kleinert} for intermediate
coupling $g=1$ for which the usual weak-coupling series would
diverge.
This theory is a systematic extension of a simple
variational approach, first developed by Feynman and Kleinert
in the path integral formalism.
Feynman introduced the path integral formalism as a
quantization regulation, that represents the operator
properties of quantum physics by fluctuations of the
dynamical variables \cite{Feynman1,Feynmanhibbs}.
By extending analytically real time to imaginary time,
also quantum statistical quantities can be obtained
by summing over quantum mechanical and thermal fluctuations
with the help of path integrals \cite{Feynmanhibbs,Feynman2}.
In order to evaluate the path integral for the free energy
approximatively, Feynman and Kleinert developed a variational
method in 1986 \cite{Feynman}. It replaces the
relevant system by the exactly solvable harmonic oscillator
whose frequency becomes a variational parameter
which has to be optimized. Starting with Ref.~\cite{Kleinert2},
this method has been systematically
extended by Kleinert to higher orders \cite{Kleinert,Frohlinde}.
It is now known as variational perturbation theory
and yields results for all temperatures and all coupling strengths.\\
In Sec.~\ref{sec10} we extend this procedure to higher orders of 
the free energy and cross check the results in Sec.~\ref{sec11}.\\
\section{Path Integral Representation}
\label{sec2}
The path integral representation for the imaginary-time evolution
amplitude of a particle of mass $M$ moving in a one dimensional
potential $V(x)$ reads \cite{Kleinert}
\begin{eqnarray}
(x_b \, \hbar \beta | x_a \, 0) &=& 
\int_{x(0)=x_a}^{x(\hbar \beta)=x_b} {\cal D} x
\exp \left\{- \frac{1}{\hbar} \int_0^{\hbar \beta} d \tau
\right.
\nonumber \\
&& \left.
\left[ \frac{M}{2} \dot{x}^2(\tau) 
+ V(x(\tau)) \right] \right\} \, .
\label{timeevolutionpathintegral} 
\end{eqnarray}
For the anharmonic oscillator potential
\begin{eqnarray}
\label{AHOpot}
V(x)= \frac{M}{2} \omega^2 x^2 +g x^4
\end{eqnarray}
the imaginary-time evolution amplitude (\ref{timeevolutionpathintegral})
can be expanded in powers of the coupling constant $g$. Thus we obtain the
perturbation series
\begin{eqnarray}
\label{2B}
\lefteqn{
(x_b \, \hbar \beta | x_a \, 0) =
(x_b \, \hbar \beta | x_a \, 0)_{\omega}} \\
&& \times
\left[ 1 - \frac{g}{\hbar} \int_0^{\hbar\beta} d \tau_1 
\langle x^4 (\tau_1) \rangle_{\omega} + 
\, ... \right] \, , \nonumber
%
\end{eqnarray}
where we have introduced the harmonic imaginary-time evolution
amplitude
\begin{eqnarray}
\label{harmonic}
\lefteqn{
(x_b \, \hbar \beta | x_a \, 0)_{\omega} \equiv
\int_{x(0)=x_a}^{x(\hbar \beta)=x_b} 
{\cal D} x } \\
&& \times
\exp \Bigg\{ -\frac{1}{\hbar} \int_{0}^{\hbar \beta} d \tau
\left[ \frac{M}{2} \dot{x}^2 (\tau) + \frac{M}{2} \omega^2 x^2 (\tau) 
\right] \Bigg\} \, , \nonumber
\end{eqnarray}
and the harmonic path expectation value
for an arbitrary functional $F[x]$:
\begin{eqnarray}
\label{expectation}
\lefteqn{
\langle F[x] \rangle_{\omega}  \equiv 
\frac{1}{(x_b \, \hbar \beta | x_a \, 0)_{\omega} }
\int_{x(0)=x_a}^{x(\hbar \beta)=x_b} {\cal D} x 
\, F[x]} \\
&&
\times
\exp \left\{ - \frac{1}{\hbar} \int_0^{\hbar \beta} d \tau
\left[ \frac{M}{2} \dot{x}^2(\tau)
+ \frac{M}{2} \omega^2 x^2(\tau) \right] \right\} \, .
\nonumber
\end{eqnarray}
The latter is evaluated with the help of
the generating functional\index{generating functional}
for the harmonic oscillator, whose path integral representation reads
\begin{eqnarray}
\label{timeevolution}
&&
(x_b \, \hbar \beta | x_a \, 0)_{\omega} [j]  
= \int_{x(0)=x_a}^{x(\hbar \beta)=x_b}
{\cal D} x  
\exp \left\{ - \frac{1}{\hbar} \int_{0}^{\hbar \beta} d \tau
\right. \nonumber \\
&&
\left.
\times \left[ \frac{M}{2} \dot{x}^2 (\tau)
+ \frac{M}{2} \omega^2 x^2 (\tau) - j(\tau) x(\tau) \right]\right\} \, ,
\end{eqnarray}
leading to \cite{Kleinert}
\begin{eqnarray}
\label{timeevolution2}
&&
(x_b \, \hbar \beta | x_a \, 0)_{\omega} [j]  = 
(x_b \, \hbar \beta | x_a \, 0)_{\omega}  \\
&&
\times
\exp \left[
\frac{1}{\hbar} \int_0^{\hbar \beta} d \tau_1 
\, x_{\rm cl} (\tau_1) j(\tau_1) 
\right.
\nonumber \\ 
&&
\left.
+ \, \frac{1}{2 \hbar^2} \int_0^{\hbar \beta} d \tau_1 \int_0^{\hbar \beta} 
d \tau_2 \,\,  G^{\rm (D)} (\tau_1 , \tau_2) j(\tau_1) j(\tau_2) \right] 
\nonumber
\end{eqnarray}
with the harmonic imaginary-time evolution amplitude
\begin{eqnarray}
\label{harmonic2}
\lefteqn{
(x_b  \, \hbar \beta |  x_a \,  0)_{\omega}
 = \sqrt { \frac{M \omega}{2 \pi \hbar \sinh \hbar \beta \omega} }  } \\
&&
\exp \left\{ - \frac{M \omega}{2 \hbar \sinh \hbar \beta \omega}
[(x_a^2+x_b^2) 
\cosh \hbar \beta \omega - 2 x_a x_b] \right\} \, .
\nonumber
\end{eqnarray}
In equation (\ref{timeevolution2})
we have introduced the classical path
\begin{eqnarray}
\label{xcl}
x_{\rm cl} (\tau ) \equiv \frac{x_a \sinh(\hbar \beta - \tau ) \omega 
+ x_b \sinh \omega \tau}{\sinh \hbar \beta \omega} \, ,
\end{eqnarray}
and the Dirichlet Green's function
\begin{eqnarray}
\label{Green}
G^{\rm (D)} (\tau_1 , \tau_2) & \equiv &  
\frac{\hbar}{M \omega}
\frac{1}{\sinh \hbar \beta \omega}  
\\
&&
\left[ \theta(\tau_1 - \tau_2) \sinh (\hbar \beta - \tau_1) \omega 
\sinh \omega \tau_2 \right. \nonumber \\
&&
\left. + \, \theta(\tau_2 - \tau_1) \sinh (\hbar \beta - \tau_2)\omega
\sinh \omega \tau_1 \right] \, .
\nonumber
\end{eqnarray}
We follow Ref.~\cite{weissbach,bachmann} and evaluate harmonic
path expectation values of polynomials in $x$ 
arising from the generating functional (\ref{timeevolution2})
according to Wick's theorem.
Let us illustrate
the procedure to reduce the power of polynomials
by the example of the harmonic path expectation value
$\langle x^n(\tau_1) \, x^m(\tau_2) \rangle_{\omega}$:
\renewcommand{\labelenumi}{(\roman{enumi})}
\begin{enumerate}
\item Contracting $x (\tau_1)$ with $x^{n-1}(\tau_1)$
and $x^m (\tau_2)$ leads to
Green's functions $G^{\rm (D)} (\tau_1, \tau_1)$ and 
$G^{\rm (D)} (\tau_1, \tau_2)$
with multiplicity
$n-1$ and $m$, respectively. The rest of the polynomial
remains within the harmonic path expectation value, leading to
$\langle x^{n-2}(\tau_1) \, x^m (\tau_2) \rangle_{\omega}$ and
$\langle x^{n-1}(\tau_1) \, x^{m-1} (\tau_2) \rangle_{\omega}$.
\item If $n>1$, extract one $x(\tau_1)$ from the path expectation value
giving $x_{\rm cl}(\tau_1)$ multiplied
by $ \langle x^{n-1}(\tau_1) x^m (\tau_2) \rangle_{\omega}$.
\item Add the terms from (i) and (ii).
\item Repeat the previous steps 
until only products of path expectation values
$ \langle x (\tau_1)\rangle_{\omega}
= x_{\rm cl} (\tau_1)$ remain. 
\end{enumerate}
With the help of this procedure, we obtain to first order
\begin{eqnarray}
\label{x4cl}
\lefteqn{
\langle x^4 (\tau_1) \rangle_{\omega} 
= x^4 _{\rm cl} (\tau_1)
}
\nonumber \\
&&
+\, 6 \, x^2 _{\rm cl} (\tau_1) \, G^{\rm (D)} (\tau_1 , \tau_1) 
+ 3 \, G^{\rm (D)^2} (\tau_1, \tau_1) \, .
\end{eqnarray}
\begin{fmffile}{weissbach}
\setlength{\unitlength}{1mm}
\section{Feynman Diagrams}
\label{sec3}
These contractions can be illustrated
by Feynman diagrams\index{Feynman diagrams} with the
following rules:
A vertex represents the integration over $\tau$
\begin{eqnarray}
\label{vertex}
\parbox{5mm}{\centerline{
\begin{fmfgraph*}(5,5)
\setval
\fmfforce{0w,1h}{v1}
\fmfforce{1w,1h}{v2}
\fmfforce{0w,0h}{v3}
\fmfforce{1w,0h}{v4}
\fmfforce{1/2w,1/2h}{v5}
\fmf{plain}{v1,v4}
\fmf{plain}{v2,v3}
\fmfdot{v5}
\end{fmfgraph*}}}
\hspace*{1mm} = \hspace*{1mm} \int_0^{\hbar \beta} d \tau \, ,
\end{eqnarray}
a line denotes the Dirichlet Green's function\index{Green's function}
\begin{eqnarray}
\label{line}
\parbox{10mm}{\centerline{
\begin{fmfgraph*}(10,10)
\setval
\fmfforce{0w,1/2h}{v1}
\fmfforce{1w,1/2h}{v2}
\fmf{plain}{v1,v2}
\fmfv{decor.size=0, label=${\scriptstyle 1}$, l.dist=1mm, l.angle=-180}{v1}
\fmfv{decor.size=0, label=${\scriptstyle 2}$, l.dist=1mm, l.angle=0}{v2}
\end{fmfgraph*}}}
\hspace*{5mm} = \hspace*{1mm} G^{\rm (D)} (\tau_1, \tau_2) \, ,
\end{eqnarray}
and a cross or a ``current'' pictures a classical path
\begin{eqnarray}
\label{cross}
\parbox{10mm}{\centerline{
\begin{fmfgraph*}(10,10)
\setval
\fmfforce{0w,1/2h}{v1}
\fmfforce{1w,1/2h}{v2}
\fmf{plain}{v1,v2}
\fmfv{decor.shape=cross,decor.filled=shaded,decor.size=3thick}{v1}
\fmfv{decor.size=0, label=${\scriptstyle 1}$, l.dist=1mm, l.angle=0}{v2}
\end{fmfgraph*}}}
\hspace*{5mm} = \hspace*{1mm} x_{\rm cl} (\tau_1) \, .
\end{eqnarray}
Inserting the harmonic path expectation value (\ref{x4cl}) 
into the perturbation expansion (\ref{2B}) leads
in first order to the diagrams 
\begin{eqnarray}
\label{firstorderfeynman}
%
\lefteqn{
\int_0^{\hbar \beta} d \tau_1 \langle x^4 (\tau_1) \rangle_{\omega} 
} \\
&&= \hspace*{0.3cm}
%
\parbox{10mm}{\centerline{
\begin{fmfgraph*}(10,14)
\setval
\fmfforce{1/2w,12/14h}{v1}
\fmfforce{0w,1/2h}{v2}
\fmfforce{1/2w,1/2h}{v3}
\fmfforce{1w,1/2h}{v4}
\fmfforce{1/2w,2/14h}{v5}
\fmf{plain}{v1,v5}
\fmf{plain}{v2,v4}
\fmfdot{v3}
\fmfv{decor.shape=cross,decor.filled=shaded,decor.size=3thick}{v1}
\fmfv{decor.shape=cross,decor.filled=shaded,decor.size=3thick}{v2}
\fmfv{decor.shape=cross,decor.filled=shaded,decor.size=3thick}{v4}
\fmfv{decor.shape=cross,decor.filled=shaded,decor.size=3thick}{v5}
\end{fmfgraph*}}}
\hspace*{1mm} + 6 \hspace*{1mm}
\parbox{10mm}{\centerline{
\begin{fmfgraph*}(10,10)
\setval
\fmfforce{1/2w,1h}{v1}
\fmfforce{0w,1/2h}{v2}
\fmfforce{1/2w,1/2h}{v3}
\fmfforce{1w,1/2h}{v4}
\fmf{plain}{v2,v4}
\fmf{plain,left=1}{v3,v1,v3}
\fmfdot{v3}
\fmfv{decor.shape=cross,decor.filled=shaded,decor.size=3thick}{v2}
\fmfv{decor.shape=cross,decor.filled=shaded,decor.size=3thick}{v4}
\end{fmfgraph*}}}
\hspace*{1mm} + 3 \hspace*{1mm}
\parbox{10mm}{\centerline{
\begin{fmfgraph*}(10,10)
\setval
\fmfforce{0w,1/2h}{v1}
\fmfforce{1/2w,1/2h}{v2}
\fmfforce{1w,1/2h}{v3}
\fmf{plain,left=1}{v1,v2,v1}
\fmf{plain,left=1}{v2,v3,v2}
\fmfdot{v2}
\end{fmfgraph*}}}
\hspace*{1mm} .
\nonumber
\end{eqnarray}
We now evaluate the first-order Feynman
diagrams in (\ref{firstorderfeynman}) for finite temperatures
and arbitrary $x_a,x_b$. Thus we will get a first-order result
for the imaginary-time evolution amplitude in (\ref{2B}).
The first diagram leads to
%
\begin{eqnarray}
\label{kreuzgraf}
\lefteqn{
\parbox{10mm}{\centerline{
\begin{fmfgraph*}(10,10)
\setval
\fmfforce{1/2w,1h}{v1}
\fmfforce{0w,1/2h}{v2}
\fmfforce{1/2w,1/2h}{v3}
\fmfforce{1w,1/2h}{v4}
\fmfforce{1/2w,0h}{v5}
\fmf{plain}{v1,v5}
\fmf{plain}{v2,v4}
\fmfdot{v3}
\fmfv{decor.shape=cross,decor.filled=shaded,decor.size=3thick}{v1}
\fmfv{decor.shape=cross,decor.filled=shaded,decor.size=3thick}{v2}
\fmfv{decor.shape=cross,decor.filled=shaded,decor.size=3thick}{v4}
\fmfv{decor.shape=cross,decor.filled=shaded,decor.size=3thick}{v5}
\end{fmfgraph*}}}
\hspace*{1mm}
=
\frac{1}{32 \omega \sinh^4 \hbar \beta \omega}
\left[
(x_a^4+x_b^4)
\left(
\sinh 4 \hbar \beta \omega
\right.\right. }
\\
&& \left.\left.
{}-8 \sinh 2 \hbar \beta \omega
+12 \hbar \beta \omega
\right)
\right.
\nonumber \\
&&
{}+ (x_a^3 x_b+x_ax_b^3)
\left(
4 \sinh 3 \hbar \beta \omega
+ 36 \sinh \hbar \beta \omega
\right.
\nonumber\\
&& \left.
{}-48 \hbar \beta \omega \cosh \hbar \beta \omega
\right) \nonumber \\
&&
\left.
{}+ x_a^2 x_b^2
\left(
-36 \sinh 2 \hbar \beta \omega
+ 48 \hbar \beta \omega
+ 24 \hbar \beta \omega \cosh 2 \hbar \beta \omega
\right)
\right]
\, , \nonumber
\end{eqnarray}
%
and the second diagram reduces to
%
\begin{eqnarray}
\label{kaulquappe}
&&
\parbox{10mm}{\centerline{
\begin{fmfgraph*}(10,10)
\setval
\fmfforce{1/2w,1h}{v1}
\fmfforce{0w,1/2h}{v2}
\fmfforce{1/2w,1/2h}{v3}
\fmfforce{1w,1/2h}{v4}
\fmf{plain}{v2,v4}
\fmf{plain,left=1}{v3,v1,v3}
\fmfdot{v3}
\fmfv{decor.shape=cross,decor.filled=shaded,decor.size=3thick}{v2}
\fmfv{decor.shape=cross,decor.filled=shaded,decor.size=3thick}{v4}
\end{fmfgraph*}}}
\hspace*{1mm} = 
\frac{\hbar}{32 M \omega^2 \sinh^3 \hbar \beta \omega}
\left[
(x_a^2+x_b^2)
\left(
\sinh 3 \hbar \beta \omega
\right.\right.  \nonumber \\
&&
\left.\left.
{}+ 9 \sinh \hbar \beta \omega
- 12 \hbar \beta \omega \cosh \hbar \beta \omega
\right)
\right.  
\nonumber \\
&&
\left.
{}+ x_a x_b
\left(
- 12 \sinh 2 \hbar \beta \omega
+ 16 \hbar \beta \omega
\right.\right. 
\nonumber \\
&& \left.\left.
{}+ 8 \hbar \beta \omega \cosh 2 \hbar \beta \omega
\right)
\right]
\, ,
\end{eqnarray}
%
whereas the last diagram turns out to be
%
\begin{eqnarray}
\label{achtergraf}
&&
\parbox{10mm}{\centerline{
\begin{fmfgraph*}(10,10)
\setval
\fmfforce{0w,1/2h}{v1}
\fmfforce{1/2w,1/2h}{v2}
\fmfforce{1w,1/2h}{v3}
\fmf{plain,left=1}{v1,v2,v1}
\fmf{plain,left=1}{v2,v3,v2}
\fmfdot{v2}
\end{fmfgraph*}}}
\hspace*{1mm} =
\frac{\hbar^2}{16 M^2 \omega^3 \sinh^2 \hbar \beta \omega}
\left(
-3 \sinh 2 \hbar \beta \omega
\right. \nonumber \\
&&
\left.
{}+ 4 \hbar \beta \omega
+ 2 \hbar \beta \omega \cosh 2 \hbar \beta \omega
\right)
\, .
\end{eqnarray}
%
So all in all we get 
the following first-order result
for the imaginary-time evolution amplitude
%
\begin{widetext}
\begin{eqnarray}
\label{endergebnis}
\lefteqn{
(x_b \, \hbar \beta | x_a \, 0) =
(x_b \, \hbar \beta | x_a \, 0)_{\omega} 
}
\nonumber \\
&&
\times
\left(
1- \frac{g}{\hbar}
\left\{
\frac{\hbar^2}{M^2 \omega^3 \sinh^2 \hbar \beta \omega}
\left[
- \frac{9}{16} \sinh 2 \hbar \beta \omega
+ \frac{3}{4} \hbar \beta \omega
+ \frac{3}{8} \hbar \beta \omega \cosh 2 \hbar \beta \omega
\right]
\right.
\right. 
\nonumber \\
&&
+ \, \frac{\hbar}{M \omega^2 \sinh^3 \hbar \beta \omega}
\left[
(x_a^2+x_b^2)
\left(
\frac{3}{16} \sinh 3 \hbar \beta \omega
+ \frac{27}{16} \sinh \hbar \beta \omega
- \frac{9}{4} \hbar \beta \omega \cosh \hbar \beta \omega
\right)
\right.
\nonumber \\
&&
\left.
+ \, x_a x_b
\left(
- \frac{9}{4} \sinh 2 \hbar \beta \omega
+ 3 \hbar \beta \omega
+ \frac{3}{2} \hbar \beta \omega \cosh 2 \hbar \beta \omega
\right) \right] 
\nonumber \\
&&
+ \, \frac{1}{\omega \sinh^4 \hbar \beta \omega}
\left[
(x_a^4 + x_b^4)
\left(
\frac{1}{32} \sinh 4 \hbar \beta \omega
- \frac{1}{4} \sinh 2 \hbar \beta \omega
+ \frac{3}{8} \hbar \beta \omega
\right)
\right.
\nonumber \\
&&
+ \, (x_a^3 x_b + x_a x_b^3)
\left(
\frac{1}{8} \sinh 3 \hbar \beta \omega
+ \frac{9}{8} \sinh \hbar \beta \omega
- \frac{3}{2} \hbar \beta \omega \cosh \hbar \beta \omega
\right) 
\nonumber \\
&&
\left. \left. \left.
+ \, x_a^2 x_b^2
\left(
- \frac{9}{8} \sinh 2 \hbar \beta \omega
+ \frac{3}{2} \hbar \beta \omega
+ \frac{3}{4} \hbar \beta \omega \cosh 2 \hbar \beta \omega
\right) \right] \right\} + ... \right) \, .
\end{eqnarray}
\end{widetext}
The imaginary-time evolution amplitude thus has
the time reversal behaviour
\begin{eqnarray}
\label{timerev}
(x_b \, \hbar \beta | x_a \, 0)
= (x_a \, \hbar \beta | x_b \, 0)^{*} \, ,
\end{eqnarray}
while it is known that
the imaginary-time evolution amplitude is real for 
one-dimensional problems.\\
\section{Partial Differential Equation}
\label{sec4}
Consider the Schr\"odinger equation for the real-time evolution
amplitude
\begin{eqnarray}
\label{realschroedinger}
\lefteqn{
i \hbar \frac{\partial}{\partial t} (x_b \, t | x_a \, 0)
} \\
&&=
- \frac{\hbar^2}{2M} \frac{\partial^2}{\partial x_b^2}
(x_b \, t | x_a \, 0) + V(x_b) \, (x_b \, t | x_a \, 0) \, .
\nonumber
\end{eqnarray}
In order to get a corresponding
quantum statistical Schr\"odinger equation
we now have to change from real time to imaginary time, 
i.e.~we have to perform the Wick rotation $t \rightarrow -i \tau$.
Thus the Schr\"odinger equation (\ref{realschroedinger})
becomes
\begin{eqnarray}
\label{schroedinger2}
- \hbar \frac{\partial}{\partial \tau} (x_b \, \tau | x_a \, 0)
&=&- \frac{\hbar^2}{2M} \frac{\partial^2}{\partial x_b^2}
(x_b \, \tau | x_a \, 0) 
\nonumber \\
&&+\, V(x_b) \, (x_b \, \tau | x_a \, 0) \, .
\end{eqnarray}
For both the real and the imaginary-time evolution amplitude
the initial condition reads
\begin{eqnarray}
\label{constraints}
(x_b \, 0 | x_a \, 0) = \delta (x_b - x_a) \, .
\end{eqnarray}
Plugging the anharmonic oscillator potential (\ref{AHOpot}) into
the Schr\"odinger equation (\ref{schroedinger2}) we finally get
\begin{eqnarray}
\label{diffeq}
&&
\left\{
- \hbar \frac{\partial}{\partial \tau}
+ \frac{\hbar^2}{2 M} \frac{\partial^2}{\partial x_b^2}
- \frac{M}{2} \omega^2 x_b^2
- g x_b^4
\right\}
\nonumber \\
&&
(x_b \, \tau | x_a \, 0)
= 0 \, .
\end{eqnarray}
Making the ansatz
\begin{eqnarray}
\label{ansatz1}
(x_b \, \tau | x_a \, 0 ) =
(x_b \, \tau | x_a \, 0 )_{\omega} \, A(x_b,x_a,\tau) \, ,
\end{eqnarray}
where $(x_b \, \tau | x_a \, 0 )_{\omega}$ is the harmonic 
imaginary-time evolution amplitude (\ref{harmonic2}), we conclude from
(\ref{diffeq}) a partial differential equation for
$A(x_b,x_a,\tau)$:
\begin{eqnarray}
\label{schroedinger}
&&
\left\{ \frac{\partial}{\partial \tau} - \frac{\hbar}{2 M}
\frac{\partial^2}{\partial x_b^2} + \omega 
\frac{x_b \cosh \omega \tau -x_a}{\sinh \omega \tau}
\frac{\partial}{\partial x_b} + \frac{g}{\hbar} x_b^4
\right\} 
\nonumber \\
&&
A(x_b,x_a,\tau) = 0 \, .
\end{eqnarray}
We now choose our ansatz for
$A(x_b,x_a,\tau)$ by introducing three expansions 
in $g$, in $x_a$ and in $x_b$, respectively. 
Also we take out the factor $\sinh^{-l} \omega \tau$,
such that the ordinary differential equations
for the expansion coefficients become as simple as possible:
\begin{eqnarray}
\label{ansatz3}
A(x_b,x_a,\tau)=
\sum_{n=0}^{\infty}
\sum_{k=0}^{2n}
\sum_{l=0}^{2k}
g^n \, \frac{c_{2k | l}^{(n)} (\tau)}{\sinh^l \omega \tau}
x_a^{2k-l} x_b^l \, .
\end{eqnarray}
In order to obtain the unperturbed result $A(x_b,x_a,\tau)=1$
for $g=0$ we need $c_{0|0}^{(0)}(\tau)=1$.
The superscript $n$ in equation (\ref{ansatz3})
denotes the perturbative order, whereas $2k$ counts
the (even) powers of the various products $x_a^i x_b^j$.
The summations over the coordinates $x_a$, $x_b$ can be
truncated at $k=4n$, because we learn from Feynman diagrammatic
considerations that the diagram with the most currents $x$ in
the $n$th order looks like
\begin{eqnarray}
\label{nkreuze}
%
\parbox{10mm}{\centerline{
\begin{fmfgraph*}(10,10)
\setval
\fmfforce{1/2w,1h}{v1}
\fmfforce{0w,1/2h}{v2}
\fmfforce{1/2w,1/2h}{v3}
\fmfforce{1w,1/2h}{v4}
\fmfforce{1/2w,0h}{v5}
\fmf{plain}{v1,v5}
\fmf{plain}{v2,v4}
\fmfdot{v3}
\fmfv{decor.shape=cross,decor.filled=shaded,decor.size=3thick}{v1}
\fmfv{decor.shape=cross,decor.filled=shaded,decor.size=3thick}{v2}
\fmfv{decor.shape=cross,decor.filled=shaded,decor.size=3thick}{v4}
\fmfv{decor.shape=cross,decor.filled=shaded,decor.size=3thick}{v5}
\end{fmfgraph*}}}
\hspace*{3mm}
\parbox{10mm}{\centerline{
\begin{fmfgraph*}(10,10)
\setval
\fmfforce{1/2w,1h}{v1}
\fmfforce{0w,1/2h}{v2}
\fmfforce{1/2w,1/2h}{v3}
\fmfforce{1w,1/2h}{v4}
\fmfforce{1/2w,0h}{v5}
\fmf{plain}{v1,v5}
\fmf{plain}{v2,v4}
\fmfdot{v3}
\fmfv{decor.shape=cross,decor.filled=shaded,decor.size=3thick}{v1}
\fmfv{decor.shape=cross,decor.filled=shaded,decor.size=3thick}{v2}
\fmfv{decor.shape=cross,decor.filled=shaded,decor.size=3thick}{v4}
\fmfv{decor.shape=cross,decor.filled=shaded,decor.size=3thick}{v5}
\end{fmfgraph*}}}
\hspace*{5mm} ... \hspace*{5mm}
\parbox{10mm}{\centerline{
\begin{fmfgraph*}(10,10)
\setval
\fmfforce{1/2w,1h}{v1}
\fmfforce{0w,1/2h}{v2}
\fmfforce{1/2w,1/2h}{v3}
\fmfforce{1w,1/2h}{v4}
\fmfforce{1/2w,0h}{v5}
\fmf{plain}{v1,v5}
\fmf{plain}{v2,v4}
\fmfdot{v3}
\fmfv{decor.shape=cross,decor.filled=shaded,decor.size=3thick}{v1}
\fmfv{decor.shape=cross,decor.filled=shaded,decor.size=3thick}{v2}
\fmfv{decor.shape=cross,decor.filled=shaded,decor.size=3thick}{v4}
\fmfv{decor.shape=cross,decor.filled=shaded,decor.size=3thick}{v5}
\end{fmfgraph*}}}
\hspace*{3mm} \, .
\end{eqnarray}
Inserting the new ansatz (\ref{ansatz3})
into the Schr\"odinger equation (\ref{schroedinger})
and arranging the indices in such a way that each term
is proportional to $x_a^{2k-l}x_b^l$, we get for the different powers of 
$g$ and for $n>0$:
%
\begin{widetext}
\begin{eqnarray}
\label{horror}
\lefteqn{
\sum_{k=0}^{2n}
\sum_{l=0}^{2k}
\frac{x_a^{2k-l} x_b^l}{\sinh^l \omega \tau}
\frac{\partial c_{2k|l}^{(n)} (\tau)}{\partial \tau}
- \frac{\hbar}{2M}
\sum_{k=-1}^{2n-1}
\sum_{l=-2}^{2k-2}
(l+2)(l+1) \frac{c_{2k+2|l+2}^{(n)} (\tau)}{\sinh^{l+2} \omega \tau}
x_a^{2k-l} x_b^l} \nonumber \\
&&
- \, \omega
\sum_{k=0}^{2n}
\sum_{l=-1}^{2k-1} (l+1)
\frac{c_{2k|l+1}^{(n)} (\tau)}{\sinh^{l+2} \omega \tau}
x_a^{2k-l} x_b^l
+ \frac{1}{\hbar}
\sum_{k=2}^{2n}
\sum_{l=4}^{2k+4}
\frac{c_{2k-4|l-4}^{(n-1)} (\tau)}{\sinh^{l-4} \omega \tau}
x_a^{2k-l}x_b^l = 0 \, .
\end{eqnarray}
\end{widetext}
Thus the sums over $k$ and over $l$ collapse and we determine the
master equation for our coefficients $c_{2k|l}^{(n)} (\tau)$
\begin{eqnarray}
\label{masterD}
&&
\frac{\partial c_{2k|l}^{(n)} (\tau)}{\partial \tau} =
(l+2)(l+1) \frac{\hbar}{2M}
\frac{c_{2k+2|l+2}^{(n)}(\tau)}{\sinh^2 \omega \tau}
\nonumber \\
&&
{}+(l+1) \omega 
\frac{c_{2k|l+1}^{(n)}(\tau)}{\sinh^2 \omega \tau}
- \frac{1}{\hbar} 
c_{2k-4|l-4}^{(n-1)} (\tau) \, \sinh^4 \omega \tau \, ,
\end{eqnarray}
which is solved by
\begin{eqnarray}
\label{master}
&&
c_{2k|l}^{(n)} (\tau) =
(l+2)(l+1) \frac{\hbar}{2M} \int d \tau
\frac{c_{2k+2|l+2}^{(n)}(\tau)}{\sinh^2 \omega \tau}
\\
&&
{}+(l+1) \omega \int d \tau
\frac{c_{2k|l+1}^{(n)}(\tau)}{\sinh^2 \omega \tau} \nonumber \\
&& - \, \frac{1}{\hbar} \int d \tau
c_{2k-4|l-4}^{(n-1)} (\tau) \, \sinh^4 \omega \tau + d_{2k|l}^{(n)} \, .
\nonumber
\end{eqnarray}
Here the $d_{2k|l}^{(n)}$ denote the integration constants which are
fixed by applying the initial condition
\begin{eqnarray}
\label{constr}
\lim_{\tau \rightarrow 0} \left|
\frac{c_{2k|l}^{(n)} (\tau)}{\sinh^l \omega \tau} \right|
< \infty \, .
\end{eqnarray}
However the above master equation (\ref{masterD})
is not valid for all $k$ and $l$. Therefore we now introduce
a set of empirical rules telling us which of the coefficients
$c_{2k|l}^{(n)} (\tau)$ have to be dropped once we write down
(\ref{master}) for any order $n$:
\begin{enumerate}
\item Drop all terms containing a $c_{2k|l}^{(n)} (\tau)$
where $2k > 4n$.
\item Drop all terms containing a $c_{2k|l}^{(n)} (\tau)$ with
$l > 2k$.
\item Neglect all terms containing a $c_{2k|l}^{(n)} (\tau)$
with any negative indices $k$ and $l$.
\end{enumerate}
To convince the reader that equation (\ref{master}) together with
this procedure leads to the correct results we now reobtain
our first-order result from (\ref{endergebnis}).
To that end we set $n=1$, such that $k$ runs from 0 to 2 and $l$ from
0 to 4. Fixing $k=2$ and counting down from $l=4$ to $l=0$ we get
\begin{widetext}
\begin{eqnarray}
\label{k2l4}
c_{4|4}^{(1)} (\tau) &=& - \frac{1}{\hbar} \int d\tau
c_{0|0}^{(0)} (\tau) \sinh^4 \omega \tau + d_{4|4}^{(1)} 
= \frac{1}{\hbar \omega} 
\left( 
\frac{1}{32} \sinh 4 \omega \tau
-\frac{1}{4} \sinh 2 \omega \tau
+\frac{3}{8} \sinh \omega \tau
\right)
\, , \\
\label{k2l3}
c_{4|3}^{(1)} (\tau) &=& 4 \omega \int d\tau
\frac{c_{4|4}^{(1)} (\tau)}{\sinh^2 \omega \tau} + d_{4|3}^{(1)} 
=\frac{1}{\hbar \omega \sinh \omega \tau}
\left( 
\frac{1}{8} \sinh 3 \omega \tau
+\frac{9}{8} \sinh \omega \tau
-\frac{3}{2} \cosh \omega \tau
\right)
\, , \\
\label{k2l2}
c_{4|2}^{(1)} (\tau) &=& 3 \omega \int d\tau
\frac{c_{4|3}^{(1)} (\tau)}{\sinh^2 \omega \tau} + d_{4|2}^{(1)} 
=\frac{1}{\hbar \omega \sinh^2 \omega \tau}
\left(
-\frac{9}{8} \sinh 2 \omega \tau
+\frac{3}{2} \omega \tau
+\frac{3}{4} \omega \tau \cosh 2 \omega \tau
\right)
\, , \\
\label{k2l1}
c_{4|1}^{(1)} (\tau) &=& 2 \omega \int d\tau
\frac{c_{4|2}^{(1)} (\tau)}{\sinh^2 \omega \tau} + d_{4|1}^{(1)} 
=\frac{1}{\hbar \omega \sinh^3 \omega \tau}
\left(
\frac{1}{8} \sinh 3 \omega \tau
+\frac{9}{8} \sinh \omega \tau
-\frac{3}{2} \omega \tau \cosh \omega \tau
\right)
\, , \\
\label{k2l0}
c_{4|0}^{(1)} (\tau) &=& \omega \int d\tau
\frac{c_{4|1}^{(1)} (\tau)}{\sinh^2 \omega \tau} + d_{4|0}^{(1)} 
=\frac{1}{\hbar \omega \sinh^4 \omega \tau}
\left(
\frac{1}{32} \sinh 4 \omega \tau
-\frac{1}{4} \sinh 2 \omega \tau
+\frac{3}{8} \sinh \omega \tau
\right)
\, .
\end{eqnarray}
Correspondingly, for $k=1$ we obtain
\begin{eqnarray}
\label{k1l2}
c_{2|2}^{(1)} (\tau) &=& \frac{6 \hbar}{M} \int d\tau
\frac{c_{4|4}^{(1)} (\tau)}{\sinh^2 \omega \tau} + d_{2|2}^{(1)} 
=\frac{1}{M \omega^2 \sinh \omega \tau}
\left(
\frac{3}{16} \sinh 3 \omega \tau
+\frac{27}{16} \sinh \omega \tau
-\frac{9}{4} \omega \tau \cosh \omega \tau
\right)
\, , \\
\label{k1l1}
c_{2|1}^{(1)} (\tau) &=& \frac{3 \hbar}{M} \int d\tau
\frac{c_{4|3}^{(1)} (\tau)}{\sinh^2 \omega \tau}
+2 \omega \int d\tau
\frac{c_{2|2}^{(1)} (\tau)}{\sinh^2 (\tau)}
+ d_{2|1}^{(1)} \nonumber \\
&=&\frac{1}{M\omega^2 \sinh^2 \omega \tau}
\left(
-\frac{9}{4} \sinh 2 \omega \tau 
+ 3 \omega \tau
+\frac{3}{2} \omega \tau \cosh 2 \omega \tau
\right)
\, , \\
\label{k1l0}
c_{2|0}^{(1)} (\tau) &=& \frac{\hbar}{M} \int d\tau
\frac{c_{4|2}^{(1)} (\tau)}{\sinh^2 \omega \tau}
+ \omega \int d\tau
\frac{c_{2|1}^{(1)} (\tau)}{\sinh^2 (\tau)}
+ d_{2|0}^{(1)} \nonumber \\
&=&\frac{1}{\hbar \omega^2 \sinh^3 \omega \tau}
\left(
\frac{3}{16} \sinh 3 \omega \tau
+\frac{27}{16} \sinh \omega \tau
-\frac{9}{4} \omega \tau \cosh \omega \tau
\right)
\, . 
\end{eqnarray}
Finally for $k=0$ we get the equation
\begin{eqnarray}
\label{k0l0}
c_{0|0}^{(1)} (\tau) &=& \frac{\hbar}{M} \int d\tau
\frac{c_{2|2}^{(1)} (\tau)}{\sinh^2 \omega \tau}
+ d_{0|0}^{(1)} 
=\frac{\hbar}{M^2 \omega^3 \sinh^2 \omega \tau}
\left(
-\frac{9}{16} \sinh 2 \omega \tau
+\frac{3}{4} \omega \tau
+\frac{3}{8} \omega \tau \cosh 2 \omega \tau
\right)
\, .
\end{eqnarray}
\end{widetext}
The path of recursion which follows from this procedure is shown
in Fig.~\ref{fig1}.
\section{Exploiting the symmetries}
\label{sec5}
As seen above we already have to solve nine ordinary
differential equations for the first-order imaginary-time evolution 
amplitude. For any order $n$ the number $p$ of integrals to solve is
\begin{eqnarray}
\label{number}
p = \sum_{j=1}^{2n+1} (2j-1)=4n^2+4n+1 \, .
\end{eqnarray}
Due to the time reversal behaviour (\ref{timerev}),
the coefficients $c_{2k|l}^{(n)} (\tau)$ show a symmetry, namely:
\begin{eqnarray}
\label{sym}
\frac{c_{2k|l}^{(n)}(\tau) }{\sinh^{l} \omega \tau} 
= \frac{c_{2k|2k-l}^{(n)} (\tau)}{\sinh^{2k-l} \omega \tau} \, .
\end{eqnarray}
Exploiting the symmetry (\ref{sym}), we can cut down the number
(\ref{number})
considerably. At first sight it is reduced to
\begin{eqnarray}
\label{number2}
p' = \sum_{j=1}^{2n+1} j =2n^2+3n+1\, ,
\end{eqnarray}
so there are only six integrals left for the first order.
But we can go even further. Employing these symmetries we can
eventually change almost all recursive {\it differential} equations
into purely {\it algebraic} ones leaving only $p''=(2n+1)$ integrations.
So for the first order we are left with three integrations only, namely
with equations (\ref{k2l4}), (\ref{k1l2}), and (\ref{k0l0}). These 
coefficients $c_{4|4}^{(1)} (\tau)$, $c_{2|2}^{(1)} (\tau)$, and
$c_{0|0}^{(1)} (\tau)$ are integrated recursively. The other
coefficients can then be obtained algebraically: Once we have
$c_{4|4}^{(1)} (\tau)$ we also know $c_{4|0}^{(1)} (\tau)$ because
of the symmetry (\ref{sym}). Comparing equation (\ref{masterD}) for
$k=2, l=4$ and $k=2, l=0$ we then obtain an algebraic equation
for $c_{4|1}^{(1)} (\tau)$. The knowledge of $c_{4|1}^{(1)} (\tau)$
gives us $c_{4|3}^{(1)} (\tau)$ 
because of the symmetry (\ref{sym})
and by comparing (\ref{masterD})
this time for $k=2, l=3$ on the one hand
and $k=2, l=1$ on the other hand
we are left with an algebraic
equation for $c_{4|2}^{(1)} (\tau)$. Thus we get all the coefficients
for $k=2$ only by solving one differential equation, namely the one for
$c_{4|4}^{(1)} (\tau)$. For $k=1$ the procedure is similar, $k=0$
only generates one coefficient anyway, namely $c_{0|0}^{(1)} (\tau)$,
which still has to be solved by evaluating one integral.
The new path of recursion is shown in Fig.~\ref{fig222}.\\
So finally 
three out of the nine first-order coefficients are obtained
by integration, three more are clear for symmetry reasons and three come
from an algebraic recursion.\\
We now generalize the algebraic part of our recursion.
Consider again the symmetry property (\ref{sym}). Differentiation on 
both sides yields
\begin{eqnarray}
\label{step1}
&&
\frac{\partial c_{2k|l}^{(n)} (\tau)}{\partial \tau} = 
\frac{1}{\sinh^{2k-2l} \omega \tau} 
\frac{\partial c_{2k|2k-l}^{(n)} (\tau)}{\partial \tau} \nonumber \\
&&- 2 (k-l) \omega \cosh \omega \tau 
\frac{c_{2k|2k-l}^{(n)} (\tau)}{\sinh^{2k-2l+1} \omega \tau} \, .
\end{eqnarray}
Now we substitute for the two partial derivatives according
to equation (\ref{masterD}).
%
%
Solving for the $(l+1)$-st coefficient 
and shifting the index $l$ down by one we obtain
\begin{widetext}
\begin{eqnarray}
\label{masterrec}
c_{2k|l}^{(n)} (\tau) &=&
- \frac{(l+1) \hbar}{2M \omega}
c_{2k+2|l+1}^{(n)} (\tau)
+ \frac{c_{2k-4|l-5}^{(n-1)} (\tau)}{\hbar \omega l}
\, \sinh^6 \omega \tau
\nonumber \\
&&
+ \, \frac{(2k-l+3)(2k-l+2) \hbar}{2 M \omega l}
\frac{c_{2k+2|2k-l+3}^{(n)} (\tau)}{\sinh^{2k-2l+2} \omega \tau}
+ \frac{2k-l+2}{l} 
\frac{c_{2k|2k-l+2}^{(n)} (\tau)}{\sinh^{2k-2l+2} \omega \tau}
\nonumber \\
&&
- \, \frac{1}{\hbar \omega l}
\frac{c_{2k-4|2k-l-3}^{(n-1)} (\tau)}{\sinh^{2k-2l-4} \omega \tau}
- \frac{(2k-2l+2) \cosh \omega \tau}{l}
\frac{c_{2k|2k-l+1}^{(n)} (\tau)}{\sinh^{2k-2l+1} \omega \tau} \, ,
\end{eqnarray}
\end{widetext}
which is the algebraic recursion relation for any 
non-diagonal coefficient
$c_{2k|l}^{(n)} (\tau)$ with \mbox{$0<l \leq k$}. (The 
coefficients with $k<l<2k$ are then clear for 
symmetry reasons.) The diagonal coefficients
$c_{2k|2k}^{(n)}(\tau)$ still have to be integrated.
\section{Combined differential and algebraic equation}
\label{sec6}
We now combine the differential recursion with the algebraic one.
As only the diagonal coefficients have to be evaluated by
integrating the differential recursive equation,
we can even further simplify the solution (\ref{master}) to our
master equation (\ref{masterD}). 
We only need it for the diagonal coefficients, for which 
$l+1=2k+1$ is always greater than
$2k$. And according to our index rule (ii), coefficients of the
shape $c_{2k|2k+1}^{(n)}$ have to be neglected. We get
\begin{eqnarray}
\label{master2}
&&
c_{2k|2k}^{(n)} (\tau) =
(2k+2)(2k+1) \frac{\hbar}{2M} \int d \tau
\frac{c_{2k+2|2k+2}^{(n)}(\tau)}{\sinh^2 \omega \tau}
\nonumber \\
&& - \, \frac{1}{\hbar} \int d \tau
c_{2k-4|2k-4}^{(n-1)} (\tau) \, \sinh^4 \omega \tau + d_{2k|2k}^{(n)} \, .
\end{eqnarray}
Index rules (i) and (iii) still have to be applied,
$k$ runs from 0 to $2n$.\\
Let us quickly summarize the combined differential and algebraic recursion
relation considering the first order as an example.
Fig.~\ref{fig222} shows all first-order coefficients for the imaginary
time evolution amplitude. Each coefficient is represented by a little 
circle. Now the coefficients on the diagonal line $2k=l$ have to be
obtained by referring to equation (\ref{master2}) together with
rules (i) and (iii). These two rules tell us which of the coefficients
either from the the same order $n$ or from the previous order $n-1$ 
have to be integrated and which ones can be put to zero.\\
Once we have the diagonal coefficients $c_{2k|2k}^{(n)}(\tau)$
we can calculate the off-diagonal ones with $l\le k$ with the
help of equation (\ref{masterrec}). The coefficients with
$k<l<2k$ are then clear for symmetry reasons.\\
Using the computer algebra programme
{\tt Maple V R7} we managed to calculate seven perturbative
orders of the imaginary-time evolution amplitude which can be found
at \cite{internet}.
\section{Free Energy}
\label{sec7}
In this section we obtain perturbative results
for the partition function by integrating the diagonal elements
of our perturbative expression for the
imaginary-time evolution amplitude from the previous sections:
\begin{figure}[t!]
\centerline{\includegraphics[width=6cm]{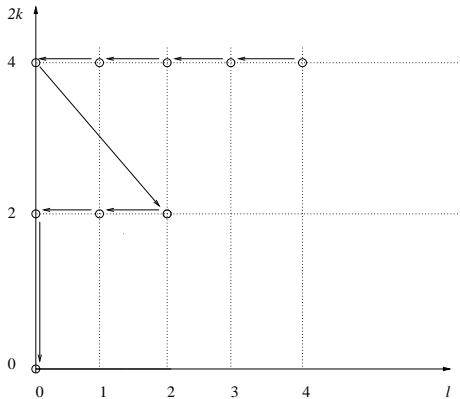}}
\caption[Path of recursion]{\label{fig1} This diagram depicts
the path of recursion for $n=1$.
We start in the top right hand side corner, which is to be identified with the
coefficient $c_{4|4}^{(1)}$ and follow the arrows until reaching
the bottom left hand side corner with the coefficient $c_{0|0}^{(1)}$.}
\end{figure}
\begin{figure}[t!]
\centerline{\includegraphics[width=6cm]{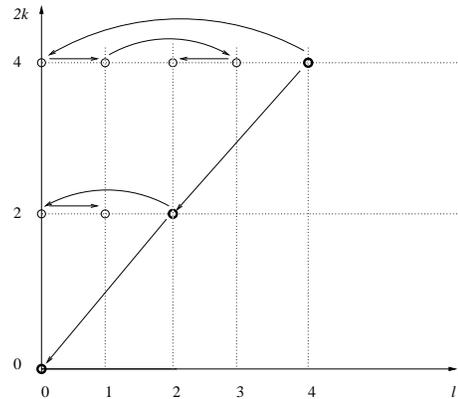}}
\caption{\label{fig222} This diagram shows which of the first-order 
coefficients $c_{2k|l}^{(1)} (\tau)$ have to be integrated (bold) and 
which ones can be obtained by employing symmetry and algebraic 
recursions (light).}
\end{figure}
\begin{eqnarray}
\label{partfctn}
Z=\int_{-\infty}^{+\infty} dx (x \, \hbar \beta | x \, 0) \, .
\end{eqnarray}
From the partition function we then compute the free energy perturbatively:
\begin{eqnarray}
\label{free}
F=- \frac{1}{\beta} \log Z \, .
\end{eqnarray}
We have to expand the logarithm in order to obtain a perturbation
expansion for the free energy $F$. 
%
%
For the first order we insert (\ref{endergebnis}) 
together with (\ref{harmonic2}) into (\ref{partfctn})
and evaluate the integral
and for the second order we use, correspondingly, the data from 
Ref.~\cite{internet}.
By taking the logarithm
we get with (\ref{free}) and with the expansion for the logarithm
for the free energy to second order
%
\begin{eqnarray}
\label{freesecond}
&&
F^{(2)}(\beta) = \frac{1}{\beta} \log 2 \sinh \frac{\hbar \beta \omega}{2} 
\\ &&
{}+ \frac{3g \hbar^2}{4 M^2 \omega^2} \coth^2 \frac{\hbar \beta \omega}{2}
-\frac{g^2 \hbar^3}{64 M^4 \omega^5}
\Bigg(\frac{54 \hbar\beta\omega}{\sinh^4 \frac{\hbar\beta\omega}{2}}
\nonumber \\
&&{}+
\frac{36 \hbar \beta \omega \cosh \hbar \beta \omega
+60 \sinh \hbar \beta \omega + 21 \sinh 2 \hbar \beta \omega}
{\sinh^4 \frac{\hbar \beta \omega}{2}} \Bigg) \, .
\nonumber
\end{eqnarray}
The higher orders are omitted for the sake of keeping the type face 
clear.\\
With {\tt Maple} we came as high as the fifth variational order
which is two orders more than what has been obtained in previous
work \cite{Meyer}.
\section{Diagrammatical Check}
\label{sec8}
It is possible to check the perturbative results
for the free energy for all temperatures. Namely, 
we can expand $Z$ in terms of harmonic
expectations in a similar way as for the imaginary-time evolution
amplitude in (\ref{2B}). To that end we need the generating
functional
\begin{eqnarray}
\label{Zj}
Z[j(\tau)]&=& \int_{-\infty}^{+\infty} dx 
(x \, \hbar \beta | x \, 0)_{\omega} [j] \, 
\end{eqnarray}
which we get from (\ref{timeevolution2})-(\ref{Green}).
It is of the form
\begin{eqnarray}
\label{Zjform}
&&Z[j(\tau)]=
Z[0] \\
&& \times
\exp \left[
\frac{1}{2 \hbar^2} \int_0^{\hbar \beta} d\tau_1 \int_0^{\hbar \beta} d\tau_2
G^{\rm (p)}(\tau_1,\tau_2) j(\tau_1) j(\tau_2) \right] \, ,
\nonumber
\end{eqnarray}
where the harmonic partition function reads
\begin{eqnarray}
\label{Z0}
Z[0]= \frac{1}{2 \sinh \frac{\hbar \beta \omega}{2}}
\end{eqnarray}
and
\begin{eqnarray}
\label{Gp}
G^{\rm (p)}(\tau_1,\tau_2)= \frac{\hbar}{2 M\omega}
\frac{\cosh \left( \frac{\hbar \beta \omega}{2}-|\tau_1-\tau_2|\omega \right)}
{\sinh \frac{\hbar \beta \omega}{2}} \, 
\end{eqnarray}
denotes the periodic Green's function of the harmonic oscillator.
We now obtain the partition function $Z$ of the anharmonic oscillator
from the generating functional
$Z[j(\tau)]$ by differentiating with respect to the current $j(\tau)$
while setting $j(\tau)=0$ afterwards:
\begin{eqnarray}
\label{Zj2}
Z = \exp \left\{ -\frac{1}{\hbar} \int_0^{\hbar \beta} d\tau \,
g \left[ \frac{\hbar \delta}{\delta j(\tau)} \right]^4 \right\}
Z[j(\tau)] \Bigg|_{j=0} \, .
\end{eqnarray}
Thus we get
\begin{eqnarray}
\label{Zexp2}
&&Z = Z[0] \left\{ 1-\frac{3g}{\hbar} \int_0^{\hbar \beta} d\tau_1
G^{\rm (p)^2}(\tau_1,\tau_1) \right.
\\
&& {}+ \frac{g^2}{2 \hbar^2} \int_0^{\hbar \beta} d\tau_1
\int_0^{\hbar \beta} d\tau_2 \left[
9 G^{\rm (p)^2}(\tau_1,\tau_1) G^{\rm (p)^2}(\tau_2,\tau_2) 
\nonumber \right. \\
&&
\left. \left.
{}+ 72 G^{\rm (p)}(\tau_1,\tau_1) G^{\rm (p)^2}(\tau_1,\tau_2) 
G^{\rm (p)}(\tau_2,\tau_2)
\right. \right.
\nonumber \\
&&
\left.\left.
{}+ 24 G^{\rm (p)^4} (\tau_1,\tau_2) \right]+... \right\} \, .
\nonumber
\end{eqnarray}
In terms of Feynman diagrams this reads
\begin{eqnarray}
\label{ZexpFeyn}
&&Z = Z[0] \left[ 1 - \frac{3g}{\hbar} \hspace*{1mm}
\parbox{10mm}{\centerline{
\begin{fmfgraph*}(10,10)
\setval
\fmfforce{0w,1/2h}{v1}
\fmfforce{1/2w,1/2h}{v2}
\fmfforce{1w,1/2h}{v3}
\fmf{plain,left=1}{v1,v2,v1}
\fmf{plain,left=1}{v2,v3,v2}
\fmfdot{v2}
\end{fmfgraph*}}}
\hspace*{1mm} 
\right. \\ && \left.
{}+ \frac{g^2}{2\hbar^2} \left( 9 \hspace*{1mm}
\parbox{10mm}{\centerline{
\begin{fmfgraph*}(10,10)
\setval
\fmfforce{0w,1/2h}{v1}
\fmfforce{1/2w,1/2h}{v2}
\fmfforce{1w,1/2h}{v3}
\fmf{plain,left=1}{v1,v2,v1}
\fmf{plain,left=1}{v2,v3,v2}
\fmfdot{v2}
\end{fmfgraph*}}}
\hspace*{1mm}
\parbox{10mm}{\centerline{
\begin{fmfgraph*}(10,10)
\setval
\fmfforce{0w,1/2h}{v1}
\fmfforce{1/2w,1/2h}{v2}
\fmfforce{1w,1/2h}{v3}
\fmf{plain,left=1}{v1,v2,v1}
\fmf{plain,left=1}{v2,v3,v2}
\fmfdot{v2}
\end{fmfgraph*}}}
\right. \right.
\nonumber \\
&&
\left. \left.
+\, 72 \hspace*{1mm}
\parbox{15mm}{\centerline{
\begin{fmfgraph*}(15,10)
\setval
\fmfforce{0w,1/2h}{v1}
\fmfforce{1/3w,1/2h}{v2}
\fmfforce{2/3w,1/2h}{v3}
\fmfforce{1w,1/2h}{v4}
\fmf{plain,left=1}{v1,v2,v1}
\fmf{plain,left=1}{v2,v3,v2}
\fmf{plain,left=1}{v3,v4,v3}
\fmfdot{v2}
\fmfdot{v3}
\end{fmfgraph*}}}
\hspace*{1mm}+ 24 \hspace*{1mm}
\parbox{10mm}{\centerline{
\begin{fmfgraph*}(10,10)
\setval
\fmfforce{0w,1/2h}{v1}
\fmfforce{1w,1/2h}{v2}
\fmf{plain,left=1}{v1,v2,v1}
\fmf{plain,left=1/2}{v1,v2}
\fmf{plain,left=1/2}{v2,v1}
\fmfdot{v1}
\fmfdot{v2}
\end{fmfgraph*}}}
\hspace*{1mm} \right)+... \right] \nonumber \\
&&=
\exp 
\left[ \frac{1}{2} \hspace*{1mm}
\parbox{5mm}{\centerline{
\begin{fmfgraph*}(5,5)
\setval
\fmfforce{1/2w,0h}{v1}
\fmfforce{1/2w,1h}{v2}
\fmf{plain,left=1}{v1,v2,v1}
\end{fmfgraph*}}}
\hspace*{1mm} - \frac{3}{\hbar} \hspace*{1mm}
\parbox{10mm}{\centerline{
\begin{fmfgraph*}(10,10)
\setval
\fmfforce{0w,1/2h}{v1}
\fmfforce{1/2w,1/2h}{v2}
\fmfforce{1w,1/2h}{v3}
\fmf{plain,left=1}{v1,v2,v1}
\fmf{plain,left=1}{v2,v3,v2}
\fmfdot{v2}
\end{fmfgraph*}}}
\hspace*{1mm} \right. \nonumber \\
&& \left.
{}+ \frac{g^2}{2 \hbar^2} \left( 72 \hspace*{1mm}
\parbox{15mm}{\centerline{
\begin{fmfgraph*}(15,10)
\setval
\fmfforce{0w,1/2h}{v1}
\fmfforce{1/3w,1/2h}{v2}
\fmfforce{2/3w,1/2h}{v3}
\fmfforce{1w,1/2h}{v4}
\fmf{plain,left=1}{v1,v2,v1}
\fmf{plain,left=1}{v2,v3,v2}
\fmf{plain,left=1}{v3,v4,v3}
\fmfdot{v2}
\fmfdot{v3}
\end{fmfgraph*}}}
\hspace*{1mm}+ 24 \hspace*{1mm}
\parbox{10mm}{\centerline{
\begin{fmfgraph*}(10,10)
\setval
\fmfforce{0w,1/2h}{v1}
\fmfforce{1w,1/2h}{v2}
\fmf{plain,left=1}{v1,v2,v1}
\fmf{plain,left=1/2}{v1,v2}
\fmf{plain,left=1/2}{v2,v1}
\fmfdot{v1}
\fmfdot{v2}
\end{fmfgraph*}}}
\hspace*{1mm} \right)+... \right] \, ,
\end{eqnarray}
where we have introduced the symbol
\begin{eqnarray}
\label{circle}
\frac{1}{2} \hspace*{1mm}
\parbox{5mm}{\centerline{
\begin{fmfgraph*}(5,5)
\setval
\fmfforce{1/2w,0h}{v1}
\fmfforce{1/2w,1h}{v2}
\fmf{plain,left=1}{v1,v2,v1}
\end{fmfgraph*}}}
\hspace*{1mm} \equiv \log Z[0] \, .
\end{eqnarray}
Once we rewrite the partition function $Z$ in the form of the cumulant
expansion as on the right hand side of equation (\ref{ZexpFeyn}),
the disconnected Feynman diagrams disappear \cite{Kleinert}.
Now we can easily take the 
logarithm. Following (\ref{free}) we obtain for
the free energy
\begin{eqnarray}
\label{FFeyn}
&&F= -\frac{1}{\beta} \left[ \frac{1}{2} \hspace*{1mm}
\parbox{5mm}{\centerline{
\begin{fmfgraph*}(5,5)
\setval
\fmfforce{1/2w,0h}{v1}
\fmfforce{1/2w,1h}{v2}
\fmf{plain,left=1}{v1,v2,v1}
\end{fmfgraph*}}}
\hspace*{1mm} - \frac{3g}{\hbar} \hspace*{1mm}
\parbox{10mm}{\centerline{
\begin{fmfgraph*}(10,10)
\setval
\fmfforce{0w,1/2h}{v1}
\fmfforce{1/2w,1/2h}{v2}
\fmfforce{1w,1/2h}{v3}
\fmf{plain,left=1}{v1,v2,v1}
\fmf{plain,left=1}{v2,v3,v2}
\fmfdot{v2}
\end{fmfgraph*}}}
\hspace*{1mm} 
\right. \\
&& \left.
{}+ \frac{g^2}{2 \hbar^2} \left( 72 \hspace*{1mm}
\parbox{15mm}{\centerline{
\begin{fmfgraph*}(15,10)
\setval
\fmfforce{0w,1/2h}{v1}
\fmfforce{1/3w,1/2h}{v2}
\fmfforce{2/3w,1/2h}{v3}
\fmfforce{1w,1/2h}{v4}
\fmf{plain,left=1}{v1,v2,v1}
\fmf{plain,left=1}{v2,v3,v2}
\fmf{plain,left=1}{v3,v4,v3}
\fmfdot{v2}
\fmfdot{v3}
\end{fmfgraph*}}}
\hspace*{1mm}+ 24 \hspace*{1mm}
\parbox{10mm}{\centerline{
\begin{fmfgraph*}(10,10)
\setval
\fmfforce{0w,1/2h}{v1}
\fmfforce{1w,1/2h}{v2}
\fmf{plain,left=1}{v1,v2,v1}
\fmf{plain,left=1/2}{v1,v2}
\fmf{plain,left=1/2}{v2,v1}
\fmfdot{v1}
\fmfdot{v2}
\end{fmfgraph*}}}
\hspace*{1mm} \right)+... \right] \, .
\nonumber
\end{eqnarray}
The above Feynman diagrams are of course constructed with the help
of the same rules as for the imaginary-time evolution amplitude
(\ref{vertex}), (\ref{line}), and (\ref{cross}), but instead of the 
Dirichlet's Green's function (\ref{Green}) we have to use the periodic Green's
function (\ref{Gp}).
We now want to evaluate the four diagrams in (\ref{FFeyn}) so that
we get a second-order expression for the free energy for finite
temperatures. According to (\ref{Z0}) and (\ref{circle}) 
we get for the zeroth-order contribution
\begin{eqnarray}
\label{Fzero}
\frac{1}{2} \hspace*{1mm}
\parbox{5mm}{\centerline{
\begin{fmfgraph*}(5,5)
\setval
\fmfforce{1/2w,0h}{v1}
\fmfforce{1/2w,1h}{v2}
\fmf{plain,left=1}{v1,v2,v1}
\end{fmfgraph*}}}
\hspace*{1mm} = \log \left[ \frac{1}{2 \sinh \frac{\hbar \beta \omega}{2}}
\right] \, ,
\end{eqnarray}
whereas the first-order diagram becomes
\begin{eqnarray}
\label{Ffirst}
\parbox{10mm}{\centerline{
\begin{fmfgraph*}(10,10)
\setval
\fmfforce{0w,1/2h}{v1}
\fmfforce{1/2w,1/2h}{v2}
\fmfforce{1w,1/2h}{v3}
\fmf{plain,left=1}{v1,v2,v1}
\fmf{plain,left=1}{v2,v3,v2}
\fmfdot{v2}
\end{fmfgraph*}}}
\hspace*{1mm} 
= \frac{\hbar^3 \beta}{4 M^2 \omega^2} \coth^2 \frac{\hbar \beta \omega}{2}
\, .
\end{eqnarray}
The integration in (\ref{Ffirst}) is trivial, because 
$G^{(\rm p)}(\tau,\tau)$
does not depend on $\tau$ any more according to (\ref{Gp}). For the second
order the integrations become more sophisticated:
\begin{eqnarray}
\label{Fsecond1}
&&
\parbox{15mm}{\centerline{
\begin{fmfgraph*}(15,10)
\setval
\fmfforce{0w,1/2h}{v1}
\fmfforce{1/3w,1/2h}{v2}
\fmfforce{2/3w,1/2h}{v3}
\fmfforce{1w,1/2h}{v4}
\fmf{plain,left=1}{v1,v2,v1}
\fmf{plain,left=1}{v2,v3,v2}
\fmf{plain,left=1}{v3,v4,v3}
\fmfdot{v2}
\fmfdot{v3}
\end{fmfgraph*}}}
\hspace*{1mm} 
= \frac{\hbar^5 \beta \coth^2 \frac{\hbar \beta \omega}{2}}
{32 M^4 \omega^5 \sinh^2 \frac{\hbar \beta \omega}{2}} 
\\
&& \times
(\hbar \beta \omega
+ \sinh \hbar \beta \omega) \, .
\nonumber
\end{eqnarray}
The other contribution to the second order yields
\begin{eqnarray}
\label{Fsecond2}
&&
\parbox{10mm}{\centerline{
\begin{fmfgraph*}(10,10)
\setval
\fmfforce{0w,1/2h}{v1}
\fmfforce{1w,1/2h}{v2}
\fmf{plain,left=1}{v1,v2,v1}
\fmf{plain,left=1/2}{v1,v2}
\fmf{plain,left=1/2}{v2,v1}
\fmfdot{v1}
\fmfdot{v2}
\end{fmfgraph*}}}
\hspace*{1mm} 
=
\frac{\hbar^5 \beta}{256 M^4 \omega^5 \sinh^4 \frac{\hbar \beta \omega}{2}}
\\
&& \times
(\sinh 2 \hbar \beta \omega
+8 \sinh \hbar \beta \omega +6 \hbar \beta \omega) \, .
\end{eqnarray}
So all in all we get for the free energy (\ref{FFeyn})
up to second order in the coupling constant $g$
the result (\ref{freesecond}).
It shows the correct low-temperature behaviour
\begin{eqnarray}
\label{limF}
\lim_{\beta \rightarrow \infty} F^{(2)}(\beta) = \frac{\hbar \omega}{2}
+ \frac{3g \hbar^2}{4 M^2 \omega^2}
- \frac{21 g^2 \hbar^3}{8 M^4 \omega^5} \, ,
\end{eqnarray}
which is the ground state energy and
can be found for instance in \cite{Bender/Wu,Kleinert}.
\section{Variational Perturbation Theory}
\label{sec9}
\begin{figure}[t!]
\centerline{\includegraphics[width=7cm]{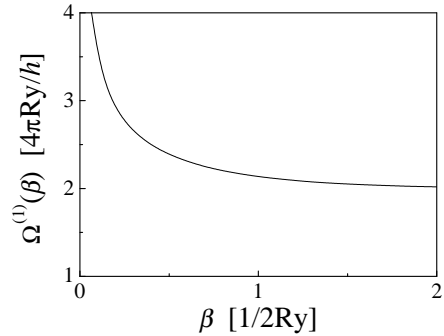}} 
\caption{\label{fig3} Branch of the variational
parameter $\Omega^{(1)} (\beta)$ which we chose. The coupling strength
is $g=1$. Other branches not shown in this figure lead to highly
diverging results. Throughout this paper 
all results are presented in natural units
$\hbar=k_{\rm B}\equiv 1$ and, additionally, we have set $M=\omega\equiv 1$.}
\end{figure}

\begin{figure}[t!]
\centerline{\includegraphics[width=7cm]{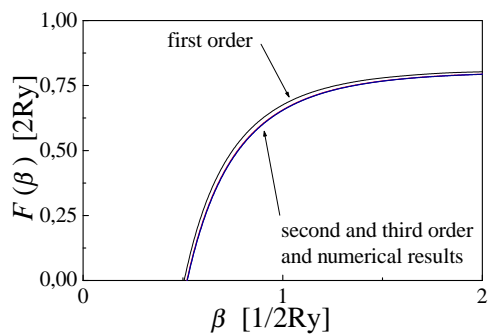}} 
\caption{\label{moin2} Free energy of the anharmonic oscillator
up to third order for intermediate coupling $g=1$.
The solid line represents the numerical result $F^{(9)}_{\rm num}(\beta)$, 
obtained by approximating the
partition function (\ref{spec2}) with the help of the first 
ten energy eigenvalues. The other lines are variational perturbative 
results: The dashed line
shows the first order, the dotted line shows the second order, 
and the dot-dashed line represents the third order. Note that the
second and third order are hardly distinguishable from the numerical results.
Higher orders for a special value of the inverse temperature $\beta$
can be found in Fig.~\ref{NELLY}.}
\end{figure}

Variational perturbation theory is a method that enables us to resum 
divergent Borel-type perturbation series in such a way that 
they converge even 
for infinitely large values of the perturbative coupling
\cite{Kleinert,Frohlinde}.
To this end we add and subtract a trial harmonic oscillator
with trial frequency $\Omega$ to our anharmonic oscillator
(\ref{AHOpot}):
\begin{eqnarray}
\label{addsubstract}
V(x)= \frac{M}{2} \Omega^2 x^2
+g \frac{M}{2} \frac{\omega^2-\Omega^2}{g} x^2 +g x^4 \, .
\end{eqnarray}
Now we treat the second term as if it was of the order 
of the coupling constant $g$. The result is obtained most simply
by substituting for the frequency $\omega$ in the original anharmonic
oscillator potential (\ref{AHOpot}) according to Kleinert's square-root
trick \cite{Kleinert}
\begin{eqnarray}
\label{trick}
\omega \rightarrow \Omega \sqrt{1+gr} \, ,
\end{eqnarray}
where
\begin{eqnarray}
\label{r}
r \equiv \frac{\omega^2-\Omega^2}{g \Omega^2} \, .
\end{eqnarray}
These substitutions are not the most general ones.
The square root is just a special case for the anharmonic
oscillator.\\
We now apply the trick (\ref{trick}) to 
our first-order series representation for
the free energy $F$ found in (\ref{freesecond}).
Substituting for the frequency $\omega$ according to
(\ref{trick}), expanding for fixed $r$ up to the first order in $g$
and re-substituting for $r$ according to (\ref{r}) we get
\begin{eqnarray}
\label{liora}
&&F^{(1)} (\beta,\Omega) = -\frac{1}{\beta} 
\log \frac{1}{2 \sinh \frac{\hbar \beta \Omega}{2}}
+  \frac{3g \hbar^2}{4 M^2 \Omega^2}
\coth^2 \frac{\hbar \beta \Omega}{2} \nonumber \\
&&
{}+\frac{\hbar \Omega}{4}
\left( \frac{\omega^2}{\Omega^2}-1 \right) \coth \frac{\hbar \beta \Omega}{2}
\, .
\end{eqnarray}
So the free energy (\ref{liora}) 
now depends on the trial frequency $\Omega$
which is of no physical relevance. In order to get rid of it,
we have to minimize its effect
by employing the principle of least sensitivity
\cite{Stevenson}.
This principle suggests to search for local extrema of 
$F(\beta,\Omega)$ with respect to $\Omega$:
\begin{eqnarray}
\label{kacke}
\frac{\partial F^{(1)}(\beta,\Omega)}{\partial \Omega} = 0 \, .
\end{eqnarray}
For the first order $F^{(1)}(\beta,\Omega)$ it turns out that
there are several extrema for each $\beta$. As we seek a curve
$\Omega^{(1)} (\beta)$ that is as smooth as possible the choice
is easy --- we take the lowest branch for the others are not
bounded (see Fig.~\ref{fig3}). Moreover the other branches
lead to diverging results.\\
To second order, we proceed in a similar way and we find that
there are no extrema at all for $F^{(2)} (\beta,\Omega)$. 
In accordance with the principle
of least sensitivity we look for inflection points instead, i.e.~we
look for solutions to the equation
\begin{eqnarray}
\label{turning}
\frac{\partial^2 F^{(2)} (\beta, \Omega)}{\partial \Omega^2} =0 \, .
\end{eqnarray}
In general we try to solve the equation
\begin{eqnarray}
\label{thumb}
\frac{\partial^n F^{(N)} (\beta, \Omega)}{\partial \Omega^n} =0 \, 
\end{eqnarray}
for the smallest possible $n$. 
Plugging $\Omega^{(N)}(\beta)$ into $F^{(N)}(\beta,\Omega)$, we finally 
get back a re-summed expression for the 
physical quantity $F(\beta)$. The results for the
first three orders are given in Fig.~\ref{moin2}.
In order to check our results we have to compare them to 
the numerically evaluated free energy $F^{(N)}_{\rm num}(\beta)$
which is discussed in Sec.~\ref{sec11}.
\section{Higher Orders}
\label{sec10}
\begin{figure}[t]
\centerline{\includegraphics[width=7cm]{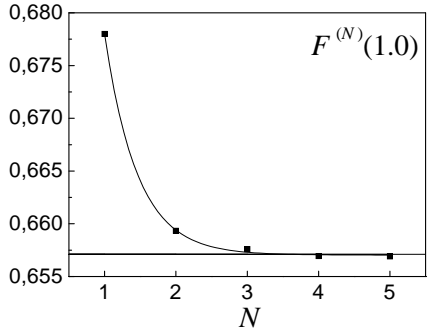}} 
\caption{\label{NELLY}
The free energy of the anharmonic oscillator
for intermediate coupling $g=1$ for $\beta=1$ 
up to fifth variational perturbative order. 
The values converge exponentially towards the numerical value
$F^{(9)}_{{\rm num}}(1)$=0.6571, which is shown as a straight line.
The dimension of the free energy in natural units is $2Ry$.}
\end{figure}

We now evaluate the convergence behaviour for the variational perturbative
results for the free energy $F^{(N)}(\beta)$ up to the fifth order.
However, in order to reduce the computational cost
we restrict ourselves to a certain value of the inverse temperature
$\beta$. Results are shown in Fig.~\ref{NELLY}.
For odd variational perturbation orders we optimized the free energy
according to (\ref{kacke}), i.e.~we determined $\Omega$ by
setting the first derivative of $F^{(N)}(\beta)$ with respect to $\Omega$ 
to zero. For even orders we had to go for inflection points, instead, 
so we had to solve equation (\ref{turning}).\\
It turns out that
odd and even orders oscillate about an exponential best fit curve.
For $\beta=1$ the numerical result, $F_{\rm num}^{(9)}(1.0)=0.6571$,
turns out to lie within the interval 
obtained by fitting the five perturbative orders of the free energy
with {\tt origin},
as shown in Fig.~\ref{NELLY}.
This interval is $[0.657,0.659]$, and
clearly, the variational perturbative results converge exponentially.
\section{Checking our Results}
\label{sec11}
The spectral representation of the partition function reads
\begin{eqnarray}
\label{spec}
Z= \sum_{n=0}^{\infty} e^{- \beta E_n} \, ,
\end{eqnarray}
where the $E_n$ are the energy eigenvalues.
Let us define the numerical approximants
\begin{eqnarray}
\label{spec2}
Z^{(N)}_{\rm num} = \sum_{n=0}^{N} e^{- \beta E_n}
\end{eqnarray}
and%
\begin{eqnarray}
\label{freeN}
F^{(N)}_{\rm num} = - \frac{1}{\beta} \log Z^{(N)} \, ,
\end{eqnarray}
respectively.
One possibility to obtain numerical results for the 
eigenvalues $E_n$ is provided for by the so called ``shooting method''.
We integrate the Schr\"odinger equation numerically for the
potential (\ref{AHOpot}) and for a particular value of the coupling strength
$g$. If the energy $E$ 
which we plug into the program does not
coincide with one of the energy eigenvalues $E_n$,
the solution to the Schr\"odinger equation explodes
already for relatively small values of the coordinate $x$.
If the energy eigenvalue is close to the exact answer, we have
$|\Psi(x)|<\infty$ also for larger values of $x$.
This method yields the unnormalized eigenfunctions (the wave functions
which still have to be normalized) and the energy eigenvalues
to very high accuracy (see Tab.~\ref{tab}). 
Renormalization is necessary, for the
computer algebra programme  needs an initial value $\Psi(0)$ which
we set to one.
\begin{table}[t!]
\centerline{
\begin{tabular}{|c|l||c|l|} \hline
$n$ & $E_n$ & $n$ & $E_n$ \\ \hline \hline
0 & 0.8037701932 & 5 & 14.203064494 \\ \hline
1 & 2.7378891484 & 6 & 17.633934116 \\ \hline
2 & 5.1792814619 & 7 & 21.236268598 \\ \hline
3 & 7.9423804544 & 8 & 24.994705012 \\ \hline
4 & 10.963538555 & 9 & 28.896941521 \\ \hline
\end{tabular}
}
\caption{\label{tab}The first ten energy eigenvalues $E_n$ of the
anharmonic oscillator for intermediate coupling $g=1$.
They were obtained by numerically integrating the Schr\"odinger
equation with the initial condition that $\Psi(0)=1$, 
$\Psi'(0)=0$ for even $n$, and $\Psi(0)=0$ and $\Psi'(0)=1$ 
for odd $n$, 
and of course $|\Psi(x)|<\infty$ for large $x$.
The energy eigenvalues are given in units of $2Ry$.}
\end{table}
Plugging the first ten numeric energy eigenvalues
\begin{figure}[t!]
\centerline{\includegraphics[width=7cm]{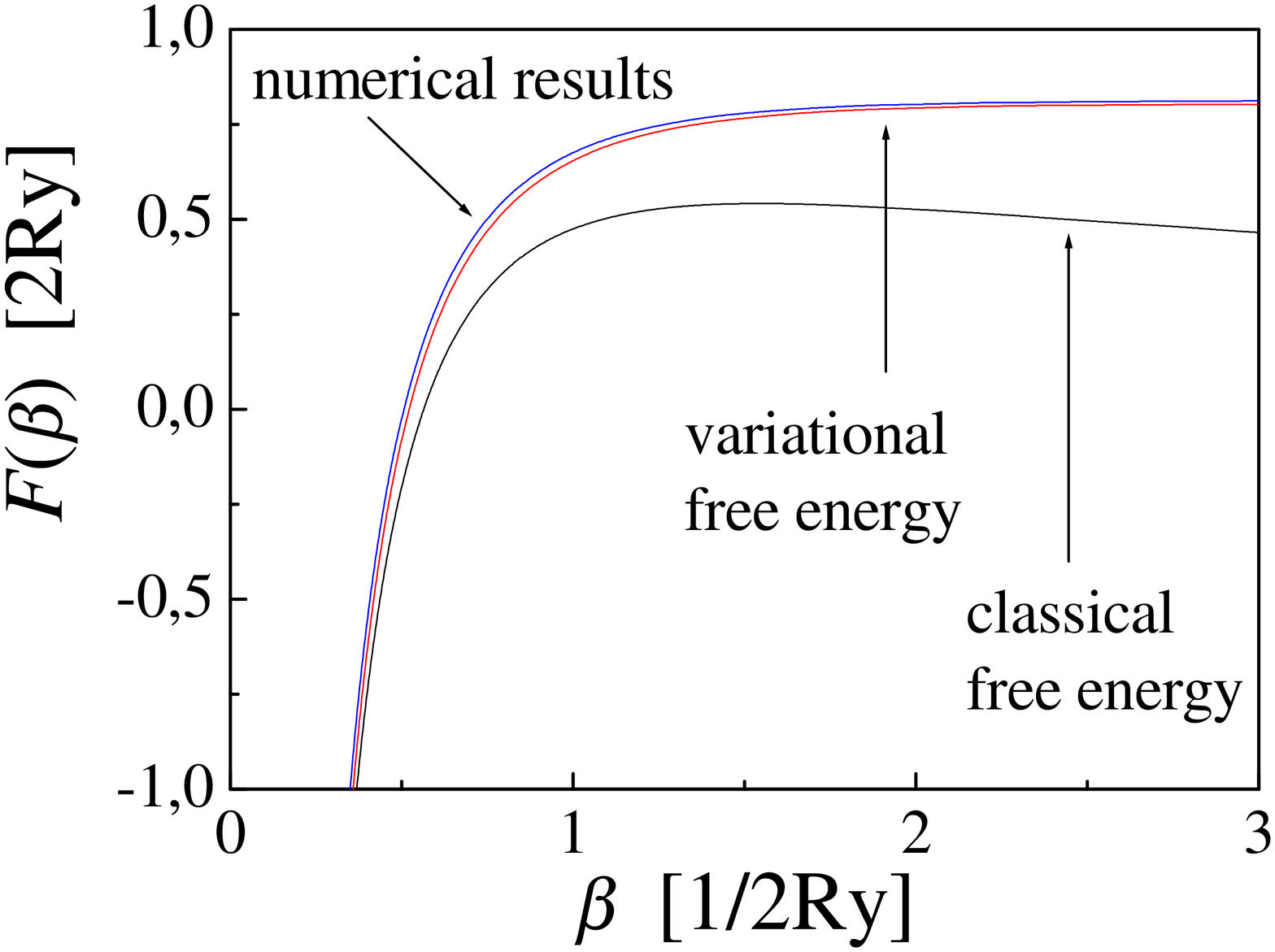}} 
\caption{\label{NELLY2} 
The numerial free energy $F_{\rm num}^{(9)}(\beta)$, the
first-order variational perturbative results for
the free energy $F^{(1)}(\beta)$,
and the classical free energy $F_{\rm cl}(\beta)=-(\log Z_{\rm cl})/\beta$,
from top to bottom.
For small values of the inverse temperature $\beta$ 
the classical calculations coincide with the other results.
Lower temperatures, corresponding to higher values of $\beta$,
reveal differences between the classical approach (\ref{NELLY1})
and quantum statistics.}
\end{figure}
into equation (\ref{spec2}) and evaluating (\ref{freeN})
up to $N=9$, we obtain a function $F^{(N)}_{{\rm num}}(\beta)$.
It converges rapidly for low temperatures, corresponding to high
values of $\beta$. For high temperatures more terms should be taken
into account.\\
Alternatively one can also use classical results as a 
high-temperature cross check:
High temperatures correspond to classical statistical distributions
such that we can evaluate the classical partition function
according to
\begin{eqnarray}
\label{clZ}
Z_{\rm cl}= \int_{-\infty}^{+ \infty} \frac{dx}{\lambda_{\rm th}}
\exp \left[ - \beta V(x) \right] \, ,
\end{eqnarray}
with the potential (\ref{AHOpot}) and
$\lambda_{\rm th}=\sqrt{2\pi \hbar^2/Mk_{\rm B}T}$.
This integral reduces to
\begin{eqnarray}
\label{NELLY1}
&&Z_{\rm cl}=\frac{1}{2 \lambda_{\rm th}} \sqrt{\frac{M\omega^2}{2g}}
\nonumber \\
&& \times
\exp \left( \frac{\beta M^2 \omega^4}{32g} \right)
K_{1/4}\left( \frac{\beta M^2 \omega^4}{32g} \right) \, ,
\end{eqnarray}
where $K_{1/4}(z)$ is a modified Bessel function.
The classical partition function (\ref{NELLY1}) can be evaluated
for high temperatures which corresponds to small values of $\beta$.
Consequently, when we test our variational perturbative results,
we compare them to the classical free energy for low values
of $\beta$, namely $\beta<1/4$. 
And for high values of $\beta$ we use the numerical
approximation, $F^{(9)}_{\rm num}(\beta)$, for comparison 
(see Fig.~\ref{NELLY2}).\\
In natural units $\hbar=k_{\rm B}=1$
a value of $\beta=1/4$ corresponds to a physical temperature
of $T=1.26 \times 10^6K$.
\section{Conclusion and Outlook}
\label{sec12}
The recursive technique that has been developed throughout Sections 
\ref{sec4}--\ref{sec6}
definitely out classes all diagrammatical 
perturbative calculations.
Using the conventional evaluation of Feynman diagrams,
the partition function and the free energy have been evaluated up to third
order \cite{Meyer}, here we obtained the fifth-order result.\\
For the free energy the convergence of variational perturbation theory 
was found to be exponential.
The fact that the principle of least sensitivity \cite{Stevenson}
produces extrema for the odd
variational orders and inflection points for even orders
is reflected in the respective convergence 
behaviours:
Odd and even orders can best be fitted separately by exponentials
as emphasized in Ref.~\cite{Frohlinde}.
Thus we obtained intervals of convergence for certain values of the free energy
which always turned out to contain the exact numerical result
when taking into account the statistical errors associated with
the boundaries of the intervals.
For the free energy, the numerical results were obtained
using its spectral representation reverting on the first ten
energy eigenvalues obtained with the ``shooting method'',
sketched in Sec.~\ref{sec11}.\\
Finally, we note that our high-order perturbative results
for the anharmonic imaginary-time evolution amplitude are useful
for calculating other thermodynamic quantities as the correlation function
or the ground state wave function \cite{weissbach,ref2,Kunihiro}.
Furthermore, it remains to compare these perturbative results
with the semiclassical approximation \cite{ref1}.
\begin{acknowledgments}
The authors wish to thank Prof.~Kleinert for fruitful discussions
on variational perturbation theory.
FW is especially grateful to Hagen Kleinert who enthusiastically
supervised his diploma thesis.
\end{acknowledgments}

\end{fmffile}
\end{document}